\documentstyle[11pt,twoside]{article}
\def\greaterthansquiggle{\raise.3ex\hbox{$>$\kern-.75em\lower1ex\hbox{$\sim$}}}
\def\lessthansquiggle{\raise.3ex\hbox{$<$\kern-.75em\lower1ex\hbox{$\sim$}}}
\newcommand{\beq}{\begin{equation}}
\newcommand{\eeq}{\end{equation}}
\newcommand{\beqa}{\begin{eqnarray}}
\newcommand{\eeqa}{\end{eqnarray}}
\newcommand{\beqan}{\begin{eqnarray*}}
\newcommand{\eeqan}{\end{eqnarray*}}
\newcommand{\ba}{\begin{array}}
\newcommand{\ea}{\end{array}}
\newcommand{\no}{\nonumber}

\newcommand{\lets}{\lessthansquiggle}

\newcommand{\ra}{\rightarrow}

\newcommand{\ve}{\varepsilon}

\newcommand{\dg}{\dagger}

\newcommand{\wh}{\widehat}

\newcommand{\cL}{{\cal L}}
\newcommand{\M}{{\cal M}}

\newcommand{\cO}{{\cal O}}

\def\nz{\ifmmode {I\hskip -3pt N} \else {\hbox {$I\hskip -3pt N$}}\fi}
\def\zz{\ifmmode {Z\hskip -4.8pt Z} \else
       {\hbox {$Z\hskip -4.8pt Z$}}\fi}
\def\qz{\ifmmode {Q\hskip -5.0pt\vrule height6.0pt depth 0pt
       \hskip 6pt} \else {\hbox
       {$Q\hskip -5.0pt\vrule height6.0pt depth 0pt\hskip 6pt$}}\fi}
\def\rz{\ifmmode {I\hskip -3pt R} \else {\hbox {$I\hskip -3pt R$}}\fi}
\def\cz{\ifmmode {C\hskip -4.8pt\vrule height5.8pt\hskip 6.3pt} \else
       {\hbox {$C\hskip -4.8pt\vrule height5.8pt\hskip 6.3pt$}}\fi}

\def\au{{\setbox0=\hbox{\lower1.36775ex%
\hbox{''}\kern-.05em}\dp0=.36775ex\hskip0pt\box0}}
\def\ao{{}\kern-.10em\hbox{``}}

\normalsize
\begin{document}
\bibliographystyle{plain}
\begin{titlepage}
\begin{flushright}
UWThPh-1994-42\\
November 17, 1994
\end{flushright}
\vspace{2cm}
\begin{center}
{\Large \bf
Isospin Violation in Chiral Perturbation Theory \\[7pt]
and the Decays $\mbox{\boldmath $\eta \ra \pi \ell \nu$}$ and
$\mbox{\boldmath $\tau \ra \eta \pi \nu$}$*}\\[50pt]
H. Neufeld \\
Institut f\"ur Theoretische Physik der Universit\"at Wien\\
Boltzmanngasse 5, A-1090 Wien, Austria\\[25pt]
A8241DAH at AWIUNI11.EDVZ.UNIVIE.AC.AT\\

\vfill
{\bf Abstract} \\
\end{center}
\noindent
I discuss isospin breaking effects within the standard
model. Chiral perturbation theory presents the appropriate
theoretical framework for such an investigation in the low--energy
range. Recent results on the electromagnetic contributions to the
masses of the pseudoscalar mesons and the $K_{\ell 3}$ amplitudes
are reported. Using the one--loop formulae for the $\eta_{\ell 3}$
form factors, rather precise predictions for the decay rates of
$\eta \rightarrow \pi \ell \nu$ can be obtained. Finally, I present
an estimate of the $\tau \ra \eta \pi \nu$ branching ratio derived
from the dominant meson resonance contributions to this decay.
\vfill
\begin{center}
To appear in the Proceedings of the
7th Adriatic Meeting in Particle Physics,\\
September 13-20, 1994, Brioni, Croatia
\vfill
\end{center}
\noindent * Supported in part by Jubil\"aumsfonds der
Oesterreichischen Nationalbank, Project No. 5051.
\end{titlepage}
\section{Introduction}
In the standard model of strong and electroweak interactions the
violation of the isospin symmetry has two different origins. First of
all, it can be traced back to the different masses of up
and down quarks. Besides that also electromagnetism
induces isospin breaking effects.

In the confinement region of the standard model the usual
perturbative methods are not applicable. In order to obtain testable
theoretical prediction also in this case one has to resort to a
so--called low--energy effective theory. With an appropriately chosen
effective Lagrangian, chiral perturbation
theory \cite{Weinberg,GL1,GL2} (which is just the effective field theory of
the standard model at low energies) is mathematically equivalent to
the underlying fundamental theory \cite{Leutwyler}. Therefore, chiral
perturbation theory presents the natural framework for the discussion
of isospin breaking effects in the low energy range.

In this report I shall first sketch the construction of the effective
chiral Lagrangian. I shall then show how electromagnetic interactions
can be incorporated into the scheme of chiral perturbation theory.
This machinery will then be applied to the mass formulae of the
pseudoscalar mesons, in particular to the question of possible
deviations from Dashen's limit \cite{Dashen}. Then I shall discuss the
$K_{\ell 3}$ and $\eta_{\ell 3}$ form factors. The latter will then
be used to obtain a standard model prediction for the decays
$\eta \rightarrow \pi \ell \nu$ ($\ell = e,\mu$). Finally we shall
leave the domain of chiral perturbation theory with an investigation
of the decay  $\tau \rightarrow \eta \pi \nu$.

This talk is based on a recent paper \cite{NeuRup} by H.
Rupertsberger and myself to which the reader is referred for a more
detailed discussion.

\section{The Low-Energy Limit of QCD}
The guiding principle for the construction of the effective
low--energy theory of QCD is the spontaneously broken chiral group
$SU(3)_L \times SU(3)_R$. The Goldstone bosons of chiral symmetry
breaking serve as the asymptotic fields of the low--energy
effective theory. With these basic building blocks one constructs
the most general relativistic quantum field theory respecting the
chiral as well as the discrete symmetries P and C.

To lowest order in the chiral expansion, the effective
Lagrangian is given by
\begin{equation}
{\cal L}_2 = \frac{F^2}{4} \, \langle D_\mu U D^\mu U^\dagger + \chi
U^\dagger + \chi^\dagger U \rangle, \label{L2}
\end{equation}
where
\begin{equation}
D_\mu U = \partial_\mu U - i (v_\mu + a_\mu) U + i U (v_\mu - a_\mu),
\quad \chi = 2
B ( s + i p ),
\end{equation}
and $\langle ... \rangle$ denotes the trace in three-dimensional
flavour space. $U$ is a unitary and unimodular $3 \times 3$ matrix
which transforms as
\begin{equation}
U \rightarrow g_R \, U  g_{L}^{\dagger}, \qquad g_{L,R} \in SU(3)_{L,R},
\end{equation}
under the chiral group. It incorporates the fields of the eight pseudoscalar
mesons.
The fields $v_\mu, a_\mu, s, p$ are  vector, axial vector,
scalar and pseudoscalar external source terms, respectively.
The quark mass matrix
\beq
\M = \mbox{diag}(m_u,m_d,m_s)
\eeq
is contained in the scalar field $s(x)$ which incorporates with $m_u
\neq m_d$ already the first source of isospin violation.
The parameters $F$
and $B$ are the only free constants of ${\cal L}_2$: $F$ is the pion decay
constant in the chiral limit, whereas $B$ is related to the quark condensate.
The Lagrangian in (\ref{L2}) is referred to as the effective chiral
Lagrangian of ${\cal O}(p^2)$. The chiral counting rules are the following:
The field $U$ is of ${\cal O}(p^0)$, the derivative $\partial_{\mu}$ and the
external gauge fields $v_{\mu},a_{\mu}$ are terms of  ${\cal O}(p)$ and the
fields $s, p$ count as  ${\cal O}(p^2)$.

At the next to leading ${\cal O}(p^4)$ the one-loop functional
generated by the lowest order Lagrangian (\ref{L2}) is renormalized by
the Gasser--Leutwyler Lagrangian \cite{GL2}
\begin{equation}
{\cal L}_4 = \sum_{i=1}^{12} L_i P_i. \label{GaLeuL}
\end{equation}
The $P_i$ are operators of ${\cal O}(p^4)$ like $P_1 =
\langle D_\mu U D^\mu U^\dagger \rangle^2$, etc. The finite parts of
the ten low--energy constants $L_1,...,L_{10}$ can be
determined \cite{GL1,GL2} by using experimental input. ($P_{11}$ and
$P_{12}$ are contact terms which are not directly accessible to experiment.)

\section{Electromagnetic Interactions}
The effective Lagrangian discussed so far allows also the description
of electromagnetic processes, provided that the electromagnetic gauge
potential $A_\mu(x)$ can be treated as an external field. As we are
interested in isospin breaking effects induced by electromagnetism,
such an approach is not sufficient anymore. In this case, the photon
field must be included as an additional dynamical degree of freedom.
One--photon loops will now contribute corrections of order $e^2$.

The corresponding local counterterms can be constructed by
introducing \cite{EGPR}
spurion fields $Q_{L,R}(x)$  which transform under the chiral
group as
\beq
Q_L \ra g_L Q_L g_L^\dg, \qquad
Q_R \ra g_R Q_R g_R^\dg.
\eeq
At the end, one identifies $Q_{L,R}$ with the quark charge matrix $Q$. The
covariant derivatives are given by
\beq
D_\mu Q_L = \partial_\mu Q_L - i[v_\mu-a_\mu,Q_L], \qquad
D_\mu Q_R = \partial_\mu Q_R - i[v_\mu+a_\mu,Q_R].
\eeq

To lowest $\cO(e^2 p^0)$ the electromagnetic effective Lagrangian
contains a single term
\beq
\left. \cL\right|_{\cO(e^2p^0)} = F^4 e^2 Z \; \langle Q U^{\dagger}
Q U\rangle,
\label{LELM}
\eeq
with a real and dimensionless coupling constant $Z$. The effective
Lagrangians (\ref{L2}) and (\ref{LELM}) contain the lowest--order
contributions to the masses of the pseudoscalar mesons from QCD and the
electromagnetic interaction, respectively:
\beqa
\wh M^2_{\pi^\pm} &=& 2B \wh m  + 2e^2 Z F^2, \no \\
\wh M^2_{\pi^0} &=& 2B \wh m , \no \\
\wh M^2_{K^\pm} &=& B\left[ (m_s + \wh m) - \frac{2\ve}{\sqrt{3}}
(m_s - \wh m)\right] + 2e^2 Z F^2, \no \\
\wh M^2_{\stackrel{(-)}{K}{}^0} &=& B \left[(m_s + \wh m) +
\frac{2\ve}{\sqrt{3}} (m_s - \wh m)\right] , \no \\
\wh M^2_\eta &=& \frac{4}{3} B\left( m_s + \frac{\wh m}{2}\right),
\label{treemass}
\eeqa
where $\wh m$ denotes the mean value of the light quark masses,
\beq
\wh m = \frac{1}{2} (m_u + m_d).
\eeq
The mixing angle
\beq
\ve = \frac{\sqrt{3}}{4} \; \frac{m_d - m_u}{m_s - \wh m} \label{epsilon}
\eeq
relates $\pi_3$, $\pi_8$ to the (tree--level) mass eigenfields
$\wh \pi_0$, $\wh \eta$:
\beqa
\pi_3 &=&  \wh \pi^0 - \ve \wh \eta, \no \\
\pi_8 &=& \ve \wh \pi^0 + \wh \eta .
\eeqa
Terms of higher than linear order in $\ve$ have been neglected. In
accordance with Dashen's theorem \cite{Dashen}, the lowest order
electromagnetic Lagrangian (\ref{LELM}) contributes an equal amount to the
squared masses of $\pi^\pm$, $K^\pm$. It does not contribute to the masses of
$\pi^0$, $K^0$, $\bar K^0$ or $\eta$, nor does it generate
$\pi^0$--$\eta$ mixing.

At next--to--leading $\cO(e^2p^2)$ one finds the following list of local
counterterms \cite{NeuRup,Urech}
\beqa
\left. \cL\right|_{\cO(e^2p^2)} &=&
F^2 e^2 \{  K_1 \; \langle Q^2\rangle \; \langle D_\mu U
D^\mu U^\dagger \rangle + K_2 \; \langle U Q U^\dagger Q \rangle \; \langle
D_\mu U D^\mu U^\dagger \rangle \no \\
&& \mbox{} + K_3 \; [\langle Q U^\dagger D_\mu U\rangle \; \langle Q
U^\dagger D^\mu U \rangle + \langle Q U D_\mu U^\dagger \rangle \;
\langle Q U D^\mu U^\dagger \rangle ] \no \\
&& \mbox{} + K_4 \; \langle D_\mu U Q U^\dagger \rangle \;
\langle D^\mu U^\dagger Q U \rangle
+ K_5 \; \langle Q^2 (D_\mu U^\dagger D^\mu U + D_\mu U
D^\mu U^\dagger \rangle \no \\
&& \mbox{} + K_6 \; \langle U Q U^\dagger Q D_\mu U D^\mu U^\dagger +
Q U Q U^\dagger D_\mu U D^\mu U^\dagger \rangle \no \\
&& \mbox{} +  K_7 \; \langle Q^2 \rangle \; \langle \chi U^\dagger
+ \chi^\dagger U \rangle
+ K_8\; \langle U Q U^\dagger Q \rangle \; \langle \chi
U^\dagger + \chi^\dagger U \rangle
\no \\
&& \mbox{}+ K_9 \; \langle Q^2 ( U^\dagger \chi + \chi^\dagger U +
\chi U^\dagger + U \chi^\dagger ) \rangle  \\
&& \mbox{} + K_{10}\; \langle Q U^\dagger Q \chi + U Q U^\dagger Q U
\chi^\dagger + U^\dagger Q U Q U^\dagger \chi + Q U Q \chi^\dagger
\rangle \no \\
&& \mbox{} - K_{11} \; \langle Q U^\dagger Q \chi - U Q U^\dagger Q U
\chi^\dagger - U^\dagger Q U Q U^\dagger \chi + Q U Q \chi^\dagger
\rangle \no \\
&& \mbox{}+ K_{12}\; \langle D_\mu U^\dagger [D^\mu Q_R,Q_R] U +
D_\mu U [D^\mu Q_L,Q_L] U^\dagger \rangle \no \\
&& \mbox{}+ K_{13} \; \langle U D_\mu Q_L U^\dagger D^\mu Q_R \rangle
+ K_{14} \; \langle D_\mu Q_L D^\mu Q_L + D_\mu Q_R D^\mu Q_R
\rangle  \}. \no \label{LE2P2}
\eeqa

As in the strong sector, the divergences of the electromagnetic parts
of the one--loop graphs are absorbed by an appropriate
renormalization \cite{Urech} of the coupling constants $K_i$.

If the class of observables is restricted to the masses of the
pseudoscalar mesons, the decay constants $F_P$ and the $P_{\ell 3}$ form
factors, we may confine ourselves to the following eight linear
combinations of the electromagnetic coupling constants \cite{NeuRup}:
\beq
\ba{lll}
S_1 = K_1 + K_2, & \qquad S_2 = K_5 + K_6, & \qquad
S_3 = -2K_3 + K_4 , \\[7pt]
S_4 = K_7 + K_8, & \qquad S_5 = K_9 + 2K_{10} + K_{11}, & \qquad
S_6 = K_8, \\[7pt]
S_7 = K_{10} + K_{11}, & \qquad S_8 = - K_{12}.
\ea
\eeq

\section{Pseudoscalar Masses}
As an illustrative example for the application of chiral
perturbation theory let me discuss some observables built out of the
masses of the pseudoscalar mesons.

The difference of the squared pion masses can be used for extracting
some information about the electromagnetic interaction. Up to tiny
corrections of $\cO(\ve^2)$,
this observable is purely electromagnetic,
\beq
M^2_{\pi^\pm} - M^2_{\pi^0} =
2e^2ZF^2 + 2e^2M_K^2 \left[ 4S_6^r(\mu) - 16 Z L_4^r(\mu) -
\frac{Z}{(4\pi)^2} \ln \frac{M_K^2}{\mu^2} \right] + \cO(e^2M_\pi^2).
\eeq
This formula exhibits the typical structure of a one--loop result in
chiral perturbation theory. It contains a so--called chiral logarithm and, in
addition, contributions from local counterterms. The scale
depence \cite{GL2,NeuRup} of the renormalized low--energy constants
is such that the total (observable) result is independent of the
scale parameter $\mu$.
At this point we are confronted with the problem that the numerical values of
the electromagnetic coupling constants $S_i^r(\mu)$
are unknown. Chiral dimensional analysis only suggests the
approximate upper bound
\beq
|S_i^r(M_\rho)| \; \lets \; \frac{1}{(4\pi)^2} = 6.3 \cdot 10^{-3}.
\label{BOUND}
\eeq

The assumption that the experimental value of $M^2_{\pi^\pm} -
M^2_{\pi^0}$ is largely dominated by the leading term of
$\cO(e^2p^0)$,
\beq
\left(M^2_{\pi^\pm} - M^2_{\pi^0} \right)_{\rm exp} \simeq
2 e^2 ZF^2, \label{PIDIFF}
\eeq
together with \cite{PDG} $F = 92.4$~MeV and \cite{GL2}
$L_4^r(M_\rho) = (- 0.3 \pm 0.5) \cdot 10^{-3}$ would imply
\beq
S_6^r(M_\rho) \approx (-2.1 \pm 1.6) \cdot 10^{-3},
\eeq
which is in accordance with (\ref{BOUND}).
Let me remark that the validity of (\ref{PIDIFF}) is also supported by
theoretical arguments \cite{EGPR,DGMLY} based on resonance exchange.

A further relation which is sensitive to electromagnetic contributions
is given by
\beqa
\frac{m^2_d - m_u^2}{m^2_s - \wh m^2} &=& \left[
(M^2_{K^0} - M^2_{K^\pm} + M^2_{\pi^\pm} - M^2_{\pi^0})_{\rm exp}
- (M^2_{K^0} - M^2_{K^\pm} + M^2_{\pi^\pm} - M^2_{\pi^0})_{\rm EM}\right] \no
\\
&& \cdot \frac{M_\pi^2}{(M_K^2 - M^2_\pi)M_K^2}, \label{mq-ratio}
\eeqa
where \cite{NeuRup,Urech}
\beqa
\label{EM}
(M^2_{K^0} - M^2_{K^\pm} + M^2_{\pi^\pm} - M^2_{\pi^0})_{\rm EM} &=&
 e^2 M_K^2 \left[ \frac{1}{(4\pi)^2}
\left( 3 \ln \frac{M_K^2}{\mu^2} - 4 + 2 Z \ln \frac{M^2_K}{\mu^2}\right)
\right. \no \\
&& \mbox{} +\left. \frac{4}{3} S_2^r(\mu) - 8 S_7^r(\mu) + 16 Z L_5^r(\mu)
\right] + \cO(e^2M_\pi^2). \no \\ \label{magicEM}
\eeqa

It is obvious that (\ref{mq-ratio}) provides us with an important
information about the quark mass ratio
\beq
1/Q^2 := \frac{m_d^2 - m_u^2}{m_s^2 - \wh m^2}. \label{defQ}
\eeq
The only uncertainty in the determination of (\ref{defQ}) is again
the unknown electromagnetic low--energy constant appearing in (\ref{magicEM}).
In Dashen's limit (corresponding
to a vanishing electromagnetic contribution),
the ratio of quark masses (\ref{defQ}) is given by
$1/Q^2 = 1.72 \cdot 10^{-3}$. This value would be reproduced for
\beq
S_2^r(M_\rho) - 6 S_7^r(M_\rho) = (24.8 \pm 4.8) \cdot 10^{-3}.
\label{numval}
\eeq
In deriving (\ref{numval}) I have used \cite{GL2}
$L_5^r(M_\rho) = (1.4 \pm 0.5) \cdot 10^{-3}$
and $Z = 0.8$ taken from (\ref{PIDIFF}). The number in (\ref{numval})
suggests again that the individual coupling constants $S_i^r(M_\rho)$ do
not exceed the rather generous order of magnitude estimate (\ref{BOUND}).
Nevertheless, because of the large coefficients occurring in (\ref{EM}),
sizeable deviations from the Dashen limit cannot be excluded.

As an illustration, let us vary the electromagnetic coupling constant
in (\ref{EM}) within the range
\beq
-\frac{7}{(4\pi)^2} \leq S_2^r(M_\rho) - 6S_7^r(M_\rho) \leq
\frac{7}{(4\pi)^2}. \label{VARIATION}
\eeq
{}From (\ref{VARIATION}) one obtains
\beq
1.5 \cdot 10^{-3} \; \lets \;
1/Q^2 = \frac{m_d^2 - m_u^2}{m_s^2 - \wh m^2}
\; \lets \; 2.4 \cdot 10^{-3}. \label{Qrange}
\eeq

The quantity $Q$ appears also in the analysis of $\eta \ra 3\pi$
decays. The calculation \cite{GL4} of these reactions has been
performed at the one--loop level in chiral perturbation theory.
As the corresponding amplitudes are proportional to $1/Q^2$, the
current experimental data \cite{PDG} can be used to extract a value
for $Q$. However, the corrections of
$\cO(p^4)$ increase the current algebra value for the transition
amplitude by roughly $50 \%$, whereas the electromagnetic
contributions are effectively of $\cO(e^2 m_{u,d})$ and thus expected
to be small \cite{GL4}. The unusually large correction of $\cO(p^4)$
is due to strong final state interaction effects. As a consequence,
the numerical accuracy of the theoretical prediction is rather
modest \cite{Leu1}. If one is nevertheless willing to use the one--loop
result for $\eta \ra 3\pi$, the present experimental data are
favouring a quark mass
ratio around $1/Q^2 = 2.3 \cdot 10^{-3}$. Values of approximately
this size are also supported by certain model calculations \cite{DHW,Bijnens}.

The knowledge of the size of the parameter Q provides us with an
important information for the determination of the quark mass ratios
$m_u/m_d$ and $m_s/m_d$. This can be seen most easily by writing
(\ref{defQ}) in the form of Leutwyler's ellipse \cite{Leumass},
\beq
(\frac{m_u}{m_d})^2 + \frac{1}{Q^2} (\frac{m_s}{m_d})^2 = 1.
\eeq
One might now worry that the large uncertainties in the
determination of $Q$ due to electromagnetism could call in question
the standard results \cite{Leumass} for the quark mass ratios.
However, there are still additional constraints \cite{Leumass} on
$m_u/m_d$ and $m_s/m_d$ from the mass splitting of the
baryons \cite{Gasser,GL6} and from an analysis of $\eta-\eta'$
mixing \cite{GL2}. The whole situation is displayed in Fig~1. We see that
in the relevant region (determined by the restrictions from $R$ and the
$\eta-\eta'$ mixing angle) the effect of a possibly large
deviation of $Q$ from Dashen's value is not too dramatic.
\begin{figure}
\setlength{\unitlength}{1mm}
\begin{picture}(90,90)(-30,0)
\put(10,10){\vector(1,0){60}}
\put(10,10){\vector(0,1){60}}
\multiput(10,10)(5,0){10}{\line(0,1){1}}
\multiput(35,10)(25,0){2}{\line(0,1){2}}
\multiput(10,10)(0,2){26}{\line(1,0){1}}
\multiput(10,20)(0,10){5}{\line(1,0){2}}
\put(60,5){\makebox(0,0){1.0}}
\put(35,5){\makebox(0,0){0.5}}
\put(10,5){\makebox(0,0){0.0}}
\put(80,10){\makebox(0,0){$m_u/m_d$}}
\put(10,75){\makebox(0,0){$m_s/m_d$}}
\put(48,56){\makebox(0,0)[l]{$|\Theta_{\eta \eta'}| = (22 \pm 4)^{\circ}$}}
\put(29,65){\makebox(0,0)[l]{$R = \frac{m_s - \hat m}{m_d - m_u} =
43.5 \pm 3.2$}}
\put(5,10){\makebox(0,0){0}}
\put(5,20){\makebox(0,0){5}}
\put(5,30){\makebox(0,0){10}}
\put(5,40){\makebox(0,0){15}}
\put(5,50){\makebox(0,0){20}}
\put(5,60){\makebox(0,0){25}}
\end{picture}
\caption{The elliptic band corresponds to the range (24).}
\end{figure}
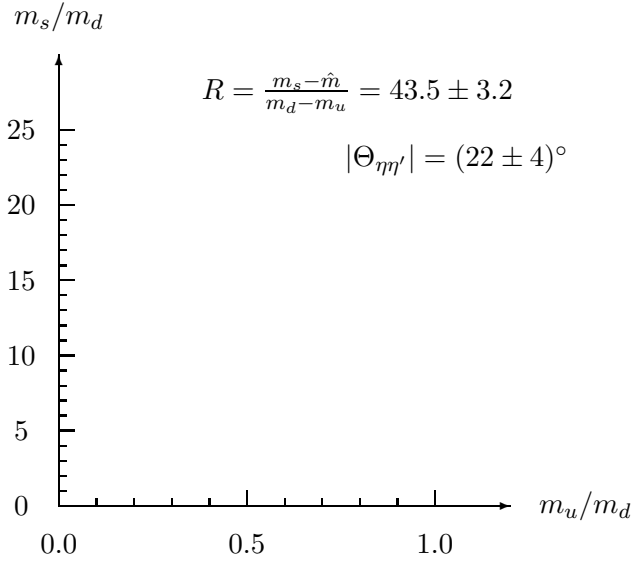

\section{$K_{\ell 3}$ form factors}
After having seen the potentially large electromagnetic contributions
in the spectrum of the pseudoscalars one might expect the same
feature also for other isospin violating observables. The following
example will show you that this is in general not the case.
For the ratio $r_{K\pi}$ of the $K_{\ell 3}$
form factors one finds the following expression \cite{NeuRup}:
\beq
r_{K\pi} = \frac{f_+^{K^+\pi^0}(0)}{f_+^{K^0 \pi^-}(0)} =
1 + \sqrt{3} \left( \ve - \frac{M^2_{\hat \pi^0 \hat \eta}}
{M_\eta^2 - M_\pi^2} \right) + \frac{3e^2}{4(4\pi)^2} \ln
\frac{M_K^2}{M_\pi^2}.
\label{RKPI}
\eeq
$M^2_{\hat \pi^0 \hat \eta}$ is the off--diagonal element of the $\pi^0 -
\eta$ mass matrix in the basis of the tree--level mass eigenfields
$\wh \pi^0, \wh \eta$. Explicitly, it is given by \cite{NeuRup}
\beqa
M^2_{\hat \pi^0 \hat \eta} &=& \frac{2\ve}{3(4\pi)^2 F^2}
\left\{ 64(4\pi)^2 (M^2_K - M^2_\pi)^2 [3L_7 + L^r_8(\mu)]\right. \no \\
&&  \mbox{} - M^2_\eta(M^2_K - M^2_\pi) \ln \frac{M^2_\eta}{\mu^2}
- 2M^2_K(M^2_K - 2 M^2_\pi) \ln \frac{M_K^2}{\mu^2} \no \\
&& \left. \mbox{} + M_\pi^2 (M^2_K - 3 M^2_\pi) \ln \frac{M_\pi^2}{\mu^2}
- 2M_K^2(M_K^2 - M_\pi^2) \right\}  \\
&& \mbox{} - \frac{2\;\sqrt{3}\;e^2\;M_K^2}{27(4\pi)^2}
\left\{2(4\pi)^2 [2 S_2^r(\mu) + 3 S_3^r(\mu)] -
9 Z \left(\ln \frac{M_K^2}{\mu^2} -1\right) \right\}. \no
\label{MASS2}
\eeqa
In the limit $e = 0$, the formula for $r_{K \pi}$ has already been
worked out \cite{LeuRoos,GL5} some time ago. The
data on the decays $K^+ \ra \pi^0 e^+ \nu_e$ and $K^0 \ra \pi^- e^+ \nu_e$ show
clear evidence for the presence of isospin breaking. Dividing the
rates by the relevant phase space integrals (including those
electromagnetic corrections which are sensitive to the lepton
kinematics \cite{LeuRoos}) one finds \cite{GL5}
\beq
|\frac{f_+^{K^+\pi^0}(0)}{f_+^{K^0 \pi^-}(0)}|^2 = 1.057 \pm 0.019,
\label{EXPRAT}
\eeq
which implies
\beq
(r_{K \pi} - 1)_{\rm exp} = (2.8 \pm 0.9) \cdot 10^{-2}.
\label{RKPIEXP}
\eeq

For our numerical analysis \cite{NeuRup} we have used the value
for $\ve$ extracted from the
mass splitting in the baryon octet \cite{Gasser,GL6},
\beq
\ve = (1.00 \pm 0.07) \cdot 10^{-2}. \label{numeps}
\eeq
With the usual numerical values \cite{GL2} for the low--energy constants
$L_7, L_8^r$, the QCD contribution to
$r_{K \pi} - 1$ is given by
\beq
(r_{K \pi} - 1)_{\rm QCD} = 2.1 \cdot 10^{-2}.
\label{RKPIQCD}
\eeq
Assuming the validity of (\ref{BOUND}), the electromagnetic
contributions are expected to be rather small,
\beq
0 \; \lets \; (r_{K \pi} - 1)_{\rm EM} \; \lets \; 0.2 \cdot 10^{-2}.
\label{RKPIEM}
\eeq

\section{$\eta \ra \pi \ell \nu$}
Also the $\eta_{\ell 3}$ form factors $f_{\pm}^{\eta\pi}(t)$ have
been calculated \cite{NeuRup} at the one--loop level in chiral
perturbation theory including the electromagnetic contributions of
$\cO(e^2p^2)$. Quite remarkably, $f_+^{\eta\pi}(0)$ can be related to
the the ratio of $K_{\ell 3}$ form factors $r_{K\pi}$ by the
parameter free relation
\beq
f_+^{\eta\pi}(0) =
\frac{1}{\sqrt{3}}[r_{K\pi} - 1 - \frac{3\;e^2}{4(4\pi)^2}
\ln \frac{M_K^2}{M_\pi^2}].
\eeq
Inserting the experimental value for $r_{K\pi}-1$ given in
(\ref{RKPIEXP}), we obtain the prediction
\beq
f_+^{\eta\pi}(0) = (1.6 \pm 0.5) \cdot 10^{-2}.
\label{fplusexp}
\eeq

Although interesting in principle, this approach is drastically
depreciated by the large error in (\ref{fplusexp}). The more promising
strategy is certainly to employ again the input parameters which have
already been used in deriving (\ref{RKPIQCD}) and (\ref{RKPIEM}). In
this case, we obtain
\beq
f_+^{\eta\pi}(0)|_{\rm QCD} = 1.21 \cdot 10^{-2}
\eeq
for the QCD contributions. As in the $K_{\ell 3}$ case the
electromagnetic contributions are expected to be rather small,
\beq
0.02 \cdot 10^{-2} \; \lets \; f_+^{\eta\pi}(0)|_{\rm EM} \;
\lets \; 0.15 \cdot
10^{-2}.
\eeq

With the full expressions for the form factors $f_{\pm}^{\eta\pi}(t)$
the branching ratios of the decays $\eta \ra \pi \ell \nu$ ($\ell =
e, \mu$) can be computed \cite{NeuRup}. Taking into account the
uncertainties due to electromagnetic contributions which have been
estimated by using (\ref{BOUND}), one obtains \cite{NeuRup}
\beqa
4.7 \cdot 10^{-14} \; \lets \; BR(\eta \ra \pi^+ e^- \bar \nu_e) \;
\lets \; 5.8 \cdot 10^{-14}, \no \\
3.4 \cdot 10^{-14} \;  \lets BR(\eta \ra \pi^+ \mu^- \bar \nu_{\mu})
\;  \lets \; 4.1 \cdot 10^{-14}.
\label{BRANCHRAT}
\eeqa
Adding all four decay channels, one arrives at \cite{NeuRup}
\beq
1.6 \cdot 10^{-13} \; \lets \;
\sum_{\ell=e,\mu} BR(\eta \ra \pi^{\pm} \ell^{\mp}
\stackrel{(-)}{\nu_\ell}) \; \lets \; 2.0 \cdot 10^{-13}.
\label{THB}
\eeq

The most powerful source of $\eta$ particles is presently installed
at the SATURNE synchrotron in Saclay. In this $\eta$ factory,
the reaction $p d \ra {}^3{\rm He} \: \eta$ serves as a source of
$\sim \! \! 10^8$ tagged $\eta$ particles per day \cite{SATURNE}.
So we see that the reaction $\eta \ra \pi \ell \nu$ is still out of
the range of present experimental facilities.
On the other hand, the observation of a decay rate
considerably larger than the upper bound in (\ref{THB}) would be
a clear signal for a deviation from the standard model.

\section{$\tau \ra \eta \pi \nu$}
The decay $\tau \ra \eta \pi \nu$ is sensitive to the same hadronic
form factors as $\eta \ra \pi \ell \nu$. However,
for the $\tau$ decay the invariant mass of the $\eta \pi$
system lies in the range $M_\eta + M_\pi \leq \sqrt{t} \leq m_\tau$, which
is already outside the domain of applicability of chiral perturbation theory.
In order to obtain reasonable theoretical results also in this
intermediate energy range the dominant contributions of the
lowest--lying resonance states have to be taken into account \cite{NeuRup}.
The presence of these resonances shows up by the
appearance of poles in the amplitudes for certain values of the
kinematical variable $t$.

Meson resonances can be incorporated in the effective chiral Lagrangian
as additional degrees of freedom \cite{EGPR,EGLPR}. They carry nonlinear
realizations of the chiral group $G$ depending on their transformation
properties under the diagonal subgroup $SU(3)_V$.

In the case of $\tau \ra \eta \pi \nu$ one has to consider the
resonance contributions from $\rho(770)$ and
$a_0(980)$. The final result \cite{NeuRup} for the hadronic form
factors  contains the vector decay constant
$F_{a_0}$ as the only free parameter. Taking
$\left| F_{a_0} \right| = 1.28 \mbox{ MeV}$
from a QCD sum rule analysis \cite{Narison}, we obtained the
prediction \cite{NeuRup}
$BR(\tau \ra \eta \pi \nu) \simeq 1.2 \cdot 10^{-5}$,
to be compared with the present experimental bound \cite{Artuso}
$BR(\tau \ra \eta \pi \nu)|_{\rm exp} < 3.4 \cdot 10^{-4}$.
Therefore, the detection of the decay $\tau \ra \eta \pi \nu$ is to
be expected in the near future. The measurement of the decay rate can
then be used for a determination of $F_{a_0}$.

\vspace{0.6cm}

\noindent {\bf Acknowledgements}
\par\vspace{0.4cm}
I wish to thank the organizers for the pleasant and stimulating
atmosphere of this conference and G. Ecker for reading the manuscript.

\newpage
\noindent  {\bf References}
\par
\end{document}